\documentclass[oldversion]{aa}
\usepackage{epsfig}
\usepackage{natbib}
\usepackage{lscape}
\usepackage{amssymb}
\usepackage{epstopdf}
\usepackage{longtable}
\usepackage{color}
\usepackage{fancyheadings}
\bibpunct{(}{)}{;}{a}{}{,}

\begin{document}
\title{Effect of stellar spots on high-precision transit light-curve}

\author{ M. Oshagh\inst{1,2} \and N. C. Santos\inst{1,2} \and I. Boisse\inst{1}  \and G. Bou{\'e}\inst{3} \and M. Montalto\inst{1}
    \and X. Dumusque \inst{4} \and N. Haghighipour\inst{5} }

\institute{
Centro de Astrof{\'\i}sica, Universidade do Porto, Rua das Estrelas, 4150-762 Porto,
Portugal \\
email: {\tt moshagh@astro.up.pt}
\and
Departamento de F{\'i}sica e Astronomia, Faculdade de Ci{\^e}ncias, Universidade do
Porto,Rua do Campo Alegre, 4169-007 Porto, Portugal
\and
Department of Astronomy and Astrophysics, University of Chicago, 5640 South Ellis Avenue, Chicago, IL, 60637, USA
\and
Observatoire de Gen\`{e}ve, Universit{\'e} de Gen\`{e}ve, 51 chemin des Maillettes, CH-1290 Sauverny, Switzerland
\and
Institute for Astronomy and NASA Astrobiology Institute, University of Hawaii-Manoa,
2680 Woodlawn Drive, Honolulu, HI 96822,USA
}

\date{Received XXX; accepted XXX}

\abstract {
Stellar-activity features such as spots can complicate the determination of
planetary parameters through spectroscopic and photometric observations. The overlap of a
transiting planet and a stellar spot, for instance, can produce anomalies in the transit light-curves that may lead to an inaccurate estimation of the transit duration, depth, and timing.
These inaccuracies can for instance affect the precise derivation of the planet radius. We present the results of a quantitative study on the effects of stellar spots on high-precision transit light-curves. We show that spot anomalies can lead to an estimate of a
planet radius that is 4\% smaller than the real value. Likewise, the transit duration may be estimated about 4\%, longer or shorter. Depending on the size and distribution of spots, anomalies can also
produce transit-timing variations (TTVs) with significant amplitudes. For instance, TTVs with signal amplitudes of 200
seconds can be produced when the spot is completely dark and has the size of the largest Sun spot. Our study also indicates that the smallest size of a stellar spot that still has detectable affects on a high-precision
transit light-curve is around 0.03 time the stellar radius for typical \emph{Kepler} telescope precision. We also
show that the strategy of including more free parameters (such as transit depth and duration) in the fitting
procedure to measure the transit time of each individual transit will not produce accurate results for
active stars.}


\authorrunning{M. Oshagh et al.}
\titlerunning{The effect of stellar spots.}
\maketitle

\section{Introduction}

A survey of the currently known extrasolar planets indicates that many of these objects orbit stars that
show high levels of activity. Among the stars in the field of view of the {\it Kepler} space telescope, for instance,
a quarter to a third are more active than the Sun \citep{Basri-13}. Active stars may also harbor spots. Indeed, spots on the stellar disk are a clear evidence of stellar activity. In general,
stellar spots are larger than Sun spots \citep{Berdyugina-05}. The largest observed stellar spots are on the giant
stars HD 12545 and II Peg \citep{ONeal-98, Strassmeier-99, Tas-00}. The spots on the surface of these stars may cover
up to half of their stellar disks.

Stellar spots can have profound effects on the detection and characterization
of planets through both photometry and spectroscopy \citep[see e.g.][]{Boisse-09,Boisse-11, Ballerini-12}. In photometric
observations, stellar spots that are not occulted by a transiting planet can produce outside-transit light-curve variations
that can lead to a wrong estimation of planet parameters such as its radius \citep{Pont-08}. The overlap of a transiting
planet and stellar spots can produce anomalies in the transit light-curve that may also lead to an incorrect
determination of planetary parameters, such as the planet radius and the limb-darkening coefficients of the host
star \citep{Pont-07, Czesla-09, Berta-11, Desert-11, Ballerini-12}. Stellar spot anomalies can also cause offsets in the transit-timing measurement that can lead to a false-positive detection of a non-transiting planet by transit-timing
variation (TTV) method \citep{Alonso-09b,Sanchis-Ojeda-11a,Sanchis-Ojeda-11b,Oshagh-12}.

To account for the anomalies generated by the overlap of a stellar spot and a transiting planet
in a fitting process, some authors assigned a zero
weight to the anomalous points of the light curve \citep[see e.g.][]{Sanchis-Ojeda-11b}. However, as shown by \citet{Barros-13}, in the presence of red noise, this treatment of stellar spot anomalies may not be the best approach since the missing
points in the transit light-curve can significantly underestimate the transit time. Moreover,
in models that are based on the symmetry of the light curve, the missing points may break the
light curve's symmetry, and as a result, the fitting routine may give more weight to the
portion of the light curve
that does not include the spot anomaly. This may cause variations in
transit timing (shifting of
the center of the light curve) and subsequently produce a TTV signal
\citep[e.g.,][]{Gibson-09,Oshagh-12}.

In this paper, we present the results of a quantitative study on the effects of stellar spots on high-precision
photometry observations. In section 2, we describe the
details of our models and initial conditions used in our simulations. Some of these require selection criteria that we also
explain in this section. In section 3, we present the results of the simulations for different cases and discuss
possible interpretations. We also determine the detectable limits on the size of a spot considering the best precision
of {\it Kepler} light curves. In section 4, we conclude our study by summarizing the results and discussing their implications.

\section{Models}

We considered a system consisting of a late spectral-type star (e.g., FGKM), and a Jupiter- or Saturn-sized
transiting planet. We chose these stars because they form more than 70\% of the stars in the solar neighborhood \citep{Henry-97},
and because they are routinely chosen as targets of transit photometry surveys (e.g., MEarth survey \citep{Charbonneau-08},
and the {\it Kepler} space telescope). These combinations of planetary sizes and late spectral-type
stars result in three values for the ratio of the planet-to-star radius, $R_{p}/R_{\ast}=$0.15, 0.1, and 0.05.
We also assumed the central star to have quadratic limb-darkening coefficients of
u1 = 0.29 and u2 = 0.34, which corresponds to a stellar temperature close to that of the Sun ($\sim$  5800 K)
\citep{Sing-10, Claret-11}. Similar coefficients have been considered for the star HD 209458
in the wavelength range of 582- 638 nm \citep{Brown-01}. In section 4.3, we
study the effect of other limb-darkening coefficients on our results.

Since we are interested in studying the maximum effect of a stellar spot, we assumed that the transiting planet
and spot completely overlap and that the spot has no brightness. This model is able to produce a light curve that is
close to a real one obtained from observations with the spot's anomaly inside the transit light-curve.
Again in section 4.3 we examine the effect of a stellar spot with non-zero brightness, as well.
For the fitting purpose, we also determined the light curve of the star excluding the effect of
the overlap between the planet and the stellar spot.

We assumed that the planet is in a three-day circular and edge-on orbit, and that the spot is located on zero latitude
corresponding to the stellar equator. We varied the filling factor of this spot defined as

\begin{equation}
f = (R_{\rm spot}/R_{\ast})^2
\end{equation}

\noindent
in the range of 0.01\% to 1\%. The upper limit of this range corresponds to the largest size of a Sun spot  \citep{Solanki-03} ($R_{\rm spot}$ and $R_{\ast}$ are the radii of the spot and star, respectively, and $f$ is defined for the visible stellar disk). The sampling time of the observation was chosen to be 60 seconds, which is equal to the short-cadence integration time of the \emph{Kepler} telescope.

\section{Simulations and results}

To study the effects of stellar spots in the light curves of our models, we used the publicly available tool
"SOAP-T"\footnote{http://www.astro.up.pt/resources/soap-t/}. The code is explained
in detail in \citet{Boisse-12} and \citet{Oshagh-12b} . This code produces the expected light curve
and the radial velocity signal of a system consisting of a rotating spotted star with a
transiting planet. SOAP-T is also able to reproduce the {\it positive bump} anomaly in the transit light-curve generated by a planet-spot overlap.

\subsection{Amplitude of anomaly}

We considered a rotating late-type star with a rotational period of nine days. We assumed that the star has a dark spot on
its surface with a filling factor of 1\%, and that it hosts a transiting planet with a radius of $R_{p}/R_{\ast}= 0.15$.
We started our simulation by considering the spot to be on the longitude and overlap with the planet while on the limb
of the star. In this case, the anomaly is produced in the egress of the transit light-curve. For the sake of comparison
and fitting purposes, we also generated a synthesis transit light-curve with the same initial conditions without considering
the overlap
between the spot and planet. The residual between the light curve of this system and the previous one where the spot and
planet overlap can be used to determine the amplitude of the anomaly. We repeated this process for different values of
the spot's longitude and determined the magnitude of the amplitude of the transit light-curve anomaly as a function of
the position of the spot on the disk of the star. As shown in Figure 1, the amplitude of the anomaly increases as the
position of the overlap between the planet and spot progresses toward the center of the stellar disk. When the overlap
occurs while the spot is on the limb of star, the amplitude of the anomaly becomes smaller than when the spot and planet
overlap at the center of the stellar disk. We note that the maximum amplitude of the anomaly is a function of the
spot's filling factor. We repeated the same analysis with a spot with a filling factor equal to 0.25 \%. As shown in Figure 1,
the amplitudes of the anomalies corresponding to the two filling factors show the same behavior as a function of the
position of the planet-spot overlap, but they have different magnitudes.

\begin{figure}
\includegraphics[width=82mm, height=50mm]{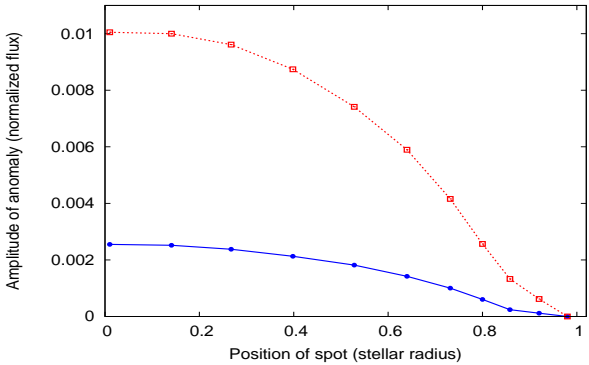}
 \caption{Amplitude of the anomaly in the transit light-curve as a function of the position
of planet-spot overlap (0 on $x$-axis means overlap occurs while the spot is at the center of stellar disk).
The values of the amplitude have been normalized to the flux value. The red dotted
line and the blue solid line correspond to transit light-curves of a transiting planet with a radius of $R_{p}/R_{\ast}= 0.15$
 and a spot with a filling a factor of 1\% and 0.25\%, respectively.}
  \label{sample-figure}
\end{figure}

\subsection{Effect of the stellar spot on the time, depth, and duration of the transit}

In this section, we examine the effect of stellar spots on the transit light-curve and its corresponding planetary parameters.
For this purpose, we generated a large number of transit light-curves for a system with a spotted star (as in section 3.1), and to understand how the spot affects the parameters of the planet, we fit these light curves with
the light curve of a system in which the effect of the spot-planet overlap was not taken into account.
In the fitting procedure, we allowed the depth, duration,
and time of the transit to vary as free parameters, keeping other parameters of the system constant to their values given
in section 3.1. The best fit of the no-anomaly light curve to each simulated transit light-curve with anomaly will give the
best value for the ratio of the planet to the stellar radius, and the time and duration of its transit. To study the effects of the
planet size and the spot filling factor, we considered the radius of the planet to be $R_{p}/R_{\ast}=0.1$ and $R_{p}/R_{\ast}=0.05$, and
generated light curves in the spot-harboring system for a zero-brightness dark spot with a filling factor of 0.25\% and 1\%.
We then fit again as explained above. Figure 2 shows some of the results: in simulations for which the position of the spot anomaly is toward the middle of the transit light-curve
(i.e., the location of the planet-spot overlap is farther from the limb and is closer to the center of the stellar disk),
the fitted value of the transit depth (that is used to derive the radius of the planet in the units of stellar radius) becomes
lower. For instance, in a system where the transiting planet has a radius of $R_{p}/R_{\ast}= 0.05$ and the spot filling
factor is 1 \%, the anomaly in the light curve causes the estimate of the planet radius to be 4\% smaller than its actual
value. This significant effect in the size-estimates of exoplanets matches the values reported for active stars and
transiting planets such as the CoRoT-2 system \citep[3\%, see][]{Czesla-09} and the system of WASP-10 (2\%), as
reported by \citet{Barros-13}.

\begin{figure}
\includegraphics[width=82mm, height=50mm]{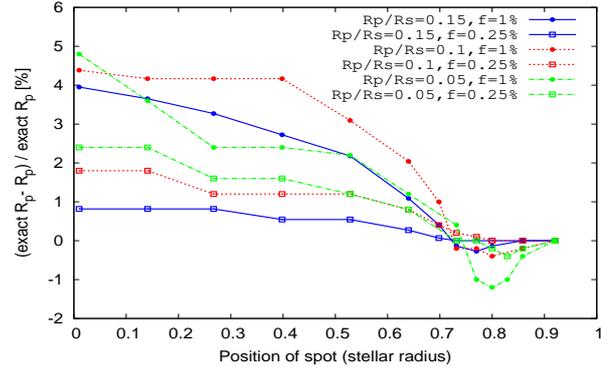}
 \caption{Deviation of the fitted value of the planet radius from its exact value as a function
of the position of planet-spot overlap. Different colors correspond to different combinations of the planet-to-star-radius ratio as well as the spot filling factor.}
  \label{sample-figure}
\end{figure}

Figure 3 shows the connection between the induced TTV and the position of the planet-spot overlap.
The TTVs were obtained by calculating the difference between the transit timing in the best-fit model and its known values.
Interestingly, the TTV amplitudes show a different behavior from that of the transit anomaly amplitude in term of
its location. As shown in the figure, significant TTVs may be produced as a result of the spot anomaly in the transit light-curve
even when the amplitude of the anomaly is not so significant. Our simulations show that the highest TTV value is reached
when the position of the overlap between the planet and the spot is at 0.7 stellar radii from the center of the star.
This result agrees with results presented by \citet{Barros-13}. This indicates that when studying the effects of
spot anomalies on variations in transit timing, one can only focus on this area where the TTV has its highest value.
Figure 3 also shows that for a transiting planet with $R_{p}/R_{\ast}= 0.1$ overlapping a spot with a filling
factor of 1 \%, the highest TTV value can exceed 200 seconds. Such a large TTV is similar to that induced by an Earth-mass
planet in a mean-motion resonance with a Jovian-type body transiting a solar-mass star in a three-day orbit \citep[e.g.,][]{Boue-12b},
or by an Earth-mass exomoon on a Neptune-mass transiting planet \citep{Kipping-09}.

\begin{figure}
\includegraphics[width=82mm, height=50mm]{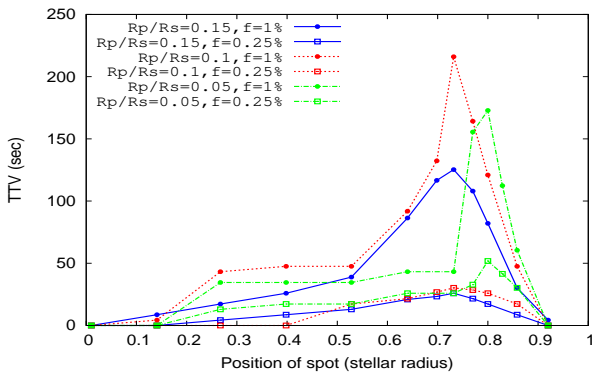}
 \caption{Amplitude of the transit-timing variations as a function of the position of
the planet-spot overlap, and for different combinations of the planet-to-star-radius ratio and the spot
filling factor.}
  \label{sample-figure}
\end{figure}

The connection between an induced spot anomaly in a transit light-curve and the duration of the transit is shown in Figure 4.
The transit duration can become shorter or longer, depending on where the anomaly appears.
The results shown here agree with those reported by \citet{Barros-13} for the WASP-10 system. Figure 4 also shows that in
extreme cases, transit durations can be under- or overestimated by about 4\%. This can be seen, for instance, for
planets with $R_{p}/R_{\ast}=0.1$ and a spot filling factor of 1\%.

\begin{figure}
\includegraphics[width=82mm, height=50mm]{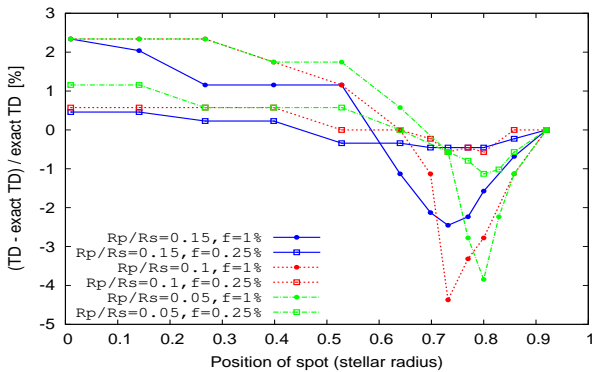}
 \caption{Deviations of the fitted value of the transit duration from its exact value as a function of
the planet-spot overlap for different combinations of the planet-to-star-radius ratio and
spot filling factor.}
  \label{sample-figure}
\end{figure}

\subsection{Probing the effect of different limb-darkening coefficients and a non-zero-brightness spot}

To probe the sensitivity of our results to the stellar spot brightness and also to quadratic limb-darkening coefficients,
we examined their influence in cases where the maximum effect on TTVs were obtained.

To assess the effect of the choice of the quadratic limb-darkening coefficients, we considered the system with the highes TTV.
In this system, the stellar spot was taken to have zero-brightness, its filling factor was 1\%, and the ratio of the radius of
the planet to that of the star was $R_{p}/R_{\ast}=0.1$. We considered two extreme cases of stellar temperatures for this system.
According to the Claret 2011 catalog \citep{Claret-11}, these extreme temperatures correspond to extreme values for the limb-darkening coefficients.
For a cool star with a temperature of 4000 K, the coefficients of the quadratic limb-darkening are u1=0.6 and u2=0.16, and
for a hot star with a temperature of 7500 K, these coefficients are u1= 0.38 and u2=0.37. Figures 5 to 7 show the effects
of these extreme quadratic limb-darkening coefficients on the depth, as well as the amplitude of the TTV and
the transit duration. Apparently, the results are not sensitive to the choice of the two extreme cases
considered here. This, however, cannot be reliable because we did not consider other possibilities. Studies similar to that of
\citet{Csizmadia-13} are needed to fully explore the effect of varying the limb-darkening coefficients on the measurements of
planetary parameters.

\begin{figure}
\includegraphics[width=82mm, height=50mm]{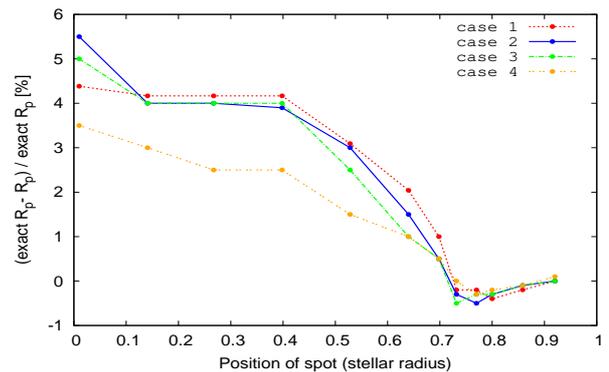}
 \caption{Deviation of the fitted value of the planet radius and its exact value as a function of
the planet-spot overlap. In cases 1 to 3, the transiting planet has a radius of $R_{p}/R_{\ast}=0.1$, the spot has zero-brightness and a filling factor of 1\%, and the limb-darkening coefficients are $({u_1},{u_2})$=(0.29,0.34),
(0.38,0.37), and (0.6,0.16), respectively. Case 4 is similar to case 1 with the brightness of the stellar spot
increased to 50\%.}
  \label{sample-figure}
\end{figure}

To study the effect of a stellar spot with a non-zero-brightness, we considered the brightness of the spot to be 50\% of that
of the star. We also increased its size by a factor 1.4 to compensate for the change in its brightness. This is equivalent
to fixing the spot filling factor to 0.1\%. As expected and similar to a zero-brightness spot, the spot with a non-zero
brightness causes variations in the time, depth, and duration of the transit, but with smaller amplitudes (Figures 5,6,7).
This is an expected result because the occultation of the non-zero-brightness spot by the transiting planet produces an anomaly that has a smaller
amplitude than that produced by a zero-brightness spot.

\begin{figure}
\includegraphics[width=82mm, height=50mm]{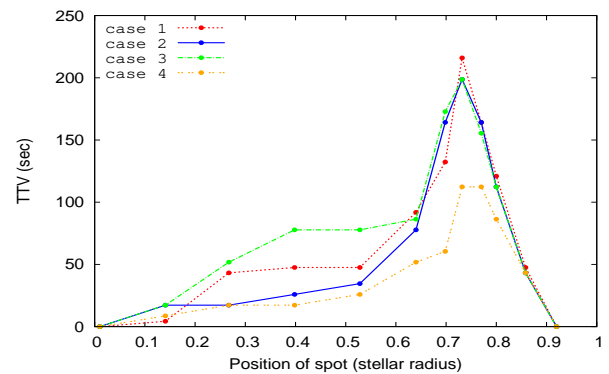}
 \caption{Amplitude of transit-timing variation as a function of the position of the planet-spot
overlap for the same cases as in figure 5.}
  \label{sample-figure}
\end{figure}

\begin{figure}
\includegraphics[width=82mm, height=50mm]{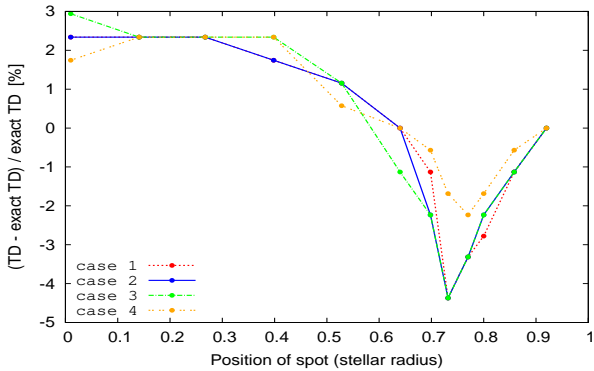}
 \caption{Deviation of the fitted value of the transit duration from its exact value as a function of the
position of the planet-spot overlap for the same cases as in figure 5.}
  \label{sample-figure}
\end{figure}

\subsection{Estimating the spot size using its induced TTV}

In this section, we estimate the maximum size of a stellar spot that will have no significant effect on a high-precision photometric observation (e.g \emph{Kepler}). The result of this study can be used to determine the minimum
detectable size of a stellar spot by using the best accessible photometric facility.

We defined the minimum size of a stellar spot as the limit for which the corresponding anomaly in the transit light-curve cannot be identified visually and will only be detectable by its effect on transit-timing measurements. We
considered a system with a transiting planet with a size of $R_{p}/R_{\ast}= 0.05$, 0.1, and 0.15 in a three-day orbit around a late-type
spotted star. We assumed that the orbit of the planet is edge-on and circular. We assigned values of $u1= 0.29$ and
$u2= 0.34$ to the star quadratic limb-darkening coefficients, and considered its rotational period to be nine days.
We placed a dark stellar spot with a zero-brightness and filling factor of 1\% on the longitude, which
corresponds to 0.7 stellar radius in the time of overlap of planet and spot. To make the simulation closer to the
real observation, we added a random Gaussian noise to each data point of the simulated light curve. The standard
deviation of the Gaussian was chosen as the best standard deviation
of a \emph{Kepler} short-cadence observation for one of its brightest target stars ($\sigma=0.00017$). We allowed the
transit timing to vary as a free parameter and fit to this system a synthetic transit light-curve that was obtained without
considering the effect of the overlap between the planet and the spot.
The TTV values were obtained from the best-fit model. We note that we only used the TTVs generated by the spot anomaly
in the transit light-curve as an indicator of the existence of a spot. Since a random noise was added to the simulation, we repeated this
process 100 times and obtained the mean value and standard deviations (shown as error bars in Figure 8) of the TTVs.
We then reduced the filling factor of the spot and determined the TTV value. Figure 8 shows the behavior of the highest TTV value as a function of spot filling factor.

To determine the minimum size of a stellar spot, we progressively reduced the spot size until the amplitude of the TTV signal
reached the \emph{Kepler} detection limit. Following \citet{Kipping-11b}, we considered this limit to be about 10 seconds.
As shown in figure 8, the minimum detectable size is obtained for a filling factor of 0.08\%, which
corresponds to a spot radius of 0.028 of the stellar radius. We would like to note that because we assumed that the overlap between planet and spot occurs at 0.7 stellar radii (where the amplitude of the effect is highest), the limit obtained from our analysis
is in fact a lower limit. Any other positions of the same spot can also produce undetectable TTV signals.

A linear fit to the highest values of TTVs obtained for different values of the planet size and stellar
spot filling factor indicates that the amplitude of TTVs caused by the stellar spot can be approximated by

\begin{equation}
AMP =
\left\{
	\begin{array}{ll}
		139 \times f  & \mbox{if } R_{p}/R_{\ast}= 0.05 \\
		132 \times f  & \mbox{if } R_{p}/R_{\ast}= 0.10 \\
        110 \times f  & \mbox{if } R_{p}/R_{\ast}= 0.15,
	\end{array}
\right.
\end{equation}

\noindent
where AMP is the maximum amplitude of TTV in units of seconds, and \emph{f} is the stellar spot filling factor in percent.

The results of our simulations show that when transit duration and depth are held constant
and equal to their known values, the TTV amplitude is smaller than when these parameters are let free.
For a transit light-curve with stellar spot anomalies in egress or ingress,
if in the fitting procedure we allow the transit duration, depth, and time to vary all as free parameters, the
chi-square of the fit becomes smaller, but will not correspond to the best fit. As shown in Figure 9, allowing
the transit duration, depth, and time to be free parameters, the best fit may pass through the anomaly by reducing the
transit duration and producing a very strong TTV signal.
However, if the transit duration and depth are held constant, contrary to what is expected that more free parameters in the fitting
process will result in more realistic planetary parameters (e.g \citealt{Mazeh-13}), and allowing for variations in the transit
timing will result in TTVs with smaller amplitudes.

\begin{figure}
\includegraphics[width=82mm, height=50mm]{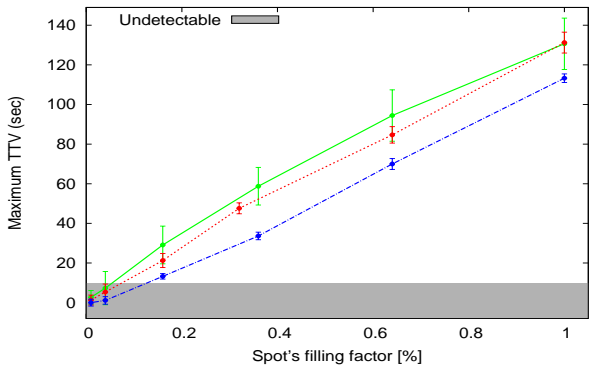}
 \caption{Highest TTV value as a function of spot filling factor. The green, dotted red, and blue dotted-dashed
lines correspond to a planet radius of $R_{p}/R_{\ast}= 0.05$, 0.1, and 0.15, respectively.}
  \label{sample-figure}
\end{figure}

\begin{figure}
\includegraphics[width=82mm, height=50mm]{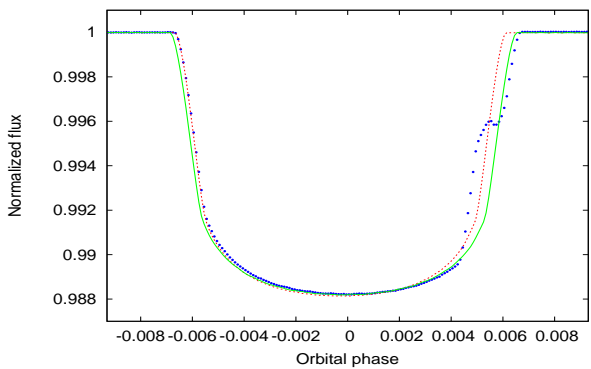}
 \caption{Transit light-curve of a star with a spot anomaly shown as a blue dot and a transiting planet with a
radius of $R_{p}/R_{\ast}= 0.1$. Overplotted is the best fit allowing transit duration, depth, and time to vary
as free parameters (red dotted line), and the best fit obtained by holding the transit duration and depth
constant and allowing only the transit timing to vary (green line).}
  \label{sample-figure}
\end{figure}

\section{Conclusions}

We presented a quantifying analysis of the effect of stellar spots on a high-precision transit light-curve.
We showed that light-curve anomalies caused by the overlap of a transiting planet and the star spot can have strong effects
on the estimate of the planet parameters. These effects can lead to an underestimation of a the planet radius by about 4\%. They can also
affect the transit duration and make it 4\% longer or shorter than its actual value. More importantly, depending on the size of the spot
and its location, an anomaly can produce transit-timing variations with an amplitudes of about 200 seconds for a typical Sun-like
spot. We also found for an active star, that allowing more free parameters (e.g., transit depth and duration) during transit fitting may
lead to an erroneous estimation of the transit timing in comparison to the case where the transit depth and duration are held constant.

Because we considered only a few cases with different values of limb-darkening coefficients, we cannot conclude
that our result are applicable to all values and combinations of these quantities. We have examined the reliability
of our result for extreme cases of limb-darkening related to extreme stellar temperatures because it is reasonable to assume that if there is going
to be any significant effect from limb-darkening, it will be from the extreme values of its coefficients. We did not, however,
find any significant effects from those values in our study.

This study can be used to properly account for the difficulties arising from stellar activities in planet characterizations
using the transit method. It also enables us to constrain the size of stellar spot using indirect techniques.

\begin{acknowledgements}

\scriptsize We acknowledge the support from the European Research Council/European Community under the
FP7 through Starting Grant agreement number 239953, and by Funda\c{c}\~ao para a Ci\^encia e a Tecnologia (FCT) in
the form of grants reference PTDC/CTE-AST/098528/2008, SFRH/BPD/81084/2011 and SFRH/BD/51981/2012. NCS also acknowledges
the support from FCT through program Ci\^encia\,2007 funded by FCT/MCTES (Portugal) and POPH/FSE (EC). NH acknowledges support
from the Hubble Space Telescope grant HST-GO-12548.06-A and from the NASA Astrobiology Institute under
Cooperative Agreement NNA09DA77 at the Institute for Astronomy, University of Hawaii.
\end{acknowledgements}

\bibliographystyle{aa}
\bibliography{mahlibspot}

\begin{thebibliography}{32}
\expandafter\ifx\csname natexlab\endcsname\relax\def\natexlab#1{#1}\fi

\bibitem[{{Alonso} {et~al.}(2009){Alonso}, {Aigrain}, {Pont}, {Mazeh}, \&
  {CoRoT Exoplanet Science Team}}]{Alonso-09b}
{Alonso}, R., {Aigrain}, S., {Pont}, F., {Mazeh}, T., \& {CoRoT Exoplanet
  Science Team}. 2009, in IAU Symposium, Vol. 253, IAU Symposium, ed.
  F.~{Pont}, D.~{Sasselov}, \& M.~J. {Holman}, 91--96

\bibitem[{{Ballerini} {et~al.}(2012){Ballerini}, {Micela}, {Lanza}, \&
  {Pagano}}]{Ballerini-12}
{Ballerini}, P., {Micela}, G., {Lanza}, A.~F., \& {Pagano}, I. 2012, \aap, 539,
  A140

\bibitem[{{Barros} {et~al.}(2013){Barros}, {Bou{\'e}}, {Gibson}, {Pollacco},
  {Santerne}, {Keenan}, {Skillen}, \& {Street}}]{Barros-13}
{Barros}, S.~C.~C., {Bou{\'e}}, G., {Gibson}, N.~P., {et~al.} 2013, \mnras,
  430, 3032

\bibitem[{{Basri} {et~al.}(2013){Basri}, {Walkowicz}, \& {Reiners}}]{Basri-13}
{Basri}, G., {Walkowicz}, L.~M., \& {Reiners}, A. 2013, \apj, 769, 37

\bibitem[{{Berdyugina}(2005)}]{Berdyugina-05}
{Berdyugina}, S.~V. 2005, Living Reviews in Solar Physics, 2, 8

\bibitem[{{Berta} {et~al.}(2011){Berta}, {Charbonneau}, {Bean}, {Irwin},
  {Burke}, {D{\'e}sert}, {Nutzman}, \& {Falco}}]{Berta-11}
{Berta}, Z.~K., {Charbonneau}, D., {Bean}, J., {et~al.} 2011, \apj, 736, 12

\bibitem[{{Boisse} {et~al.}(2012){Boisse}, {Bonfils}, \& {Santos}}]{Boisse-12}
{Boisse}, I., {Bonfils}, X., \& {Santos}, N.~C. 2012, \aap, 545, A109

\bibitem[{{Boisse} {et~al.}(2011){Boisse}, {Bouchy}, {H{\'e}brard}, {Bonfils},
  {Santos}, \& {Vauclair}}]{Boisse-11}
{Boisse}, I., {Bouchy}, F., {H{\'e}brard}, G., {et~al.} 2011, \aap, 528, A4

\bibitem[{{Boisse} {et~al.}(2009){Boisse}, {Moutou}, {Vidal-Madjar}, {Bouchy},
  {Pont}, {H{\'e}brard}, {Bonfils}, {Croll}, {Delfosse}, {Desort}, {Forveille},
  {Lagrange}, {Loeillet}, {Lovis}, {Matthews}, {Mayor}, {Pepe}, {Perrier},
  {Queloz}, {Rowe}, {Santos}, {S{\'e}gransan}, \& {Udry}}]{Boisse-09}
{Boisse}, I., {Moutou}, C., {Vidal-Madjar}, A., {et~al.} 2009, \aap, 495, 959

\bibitem[{{Bou{\'e}} {et~al.}(2012){Bou{\'e}}, {Oshagh}, {Montalto}, \&
  {Santos}}]{Boue-12b}
{Bou{\'e}}, G., {Oshagh}, M., {Montalto}, M., \& {Santos}, N.~C. 2012, \mnras,
  422, L57

\bibitem[{{Brown} {et~al.}(2001){Brown}, {Charbonneau}, {Gilliland}, {Noyes},
  \& {Burrows}}]{Brown-01}
{Brown}, T.~M., {Charbonneau}, D., {Gilliland}, R.~L., {Noyes}, R.~W., \&
  {Burrows}, A. 2001, \apj, 552, 699

\bibitem[{{Charbonneau} {et~al.}(2008){Charbonneau}, {Irwin}, {Nutzman}, \&
  {Falco}}]{Charbonneau-08}
{Charbonneau}, D., {Irwin}, J., {Nutzman}, P., \& {Falco}, E.~E. 2008, in
  Bulletin of the American Astronomical Society, Vol.~40, American Astronomical
  Society Meeting Abstracts \#212, 242

\bibitem[{{Claret} \& {Bloemen}(2011)}]{Claret-11}
{Claret}, A. \& {Bloemen}, S. 2011, VizieR Online Data Catalog, 352, 99075

\bibitem[{{Csizmadia} {et~al.}(2013){Csizmadia}, {Pasternacki}, {Dreyer},
  {Cabrera}, {Erikson}, \& {Rauer}}]{Csizmadia-13}
{Csizmadia}, S., {Pasternacki}, T., {Dreyer}, C., {et~al.} 2013, \aap, 549, A9

\bibitem[{{Czesla} {et~al.}(2009){Czesla}, {Huber}, {Wolter}, {Schr{\"o}ter},
  \& {Schmitt}}]{Czesla-09}
{Czesla}, S., {Huber}, K.~F., {Wolter}, U., {Schr{\"o}ter}, S., \& {Schmitt},
  J.~H.~M.~M. 2009, \aap, 505, 1277

\bibitem[{{D{\'e}sert} {et~al.}(2011){D{\'e}sert}, {Charbonneau}, {Demory},
  {Ballard}, {Carter}, {Fortney}, {Cochran}, {Endl}, {Quinn}, {Isaacson},
  {Fressin}, {Buchhave}, {Latham}, {Knutson}, {Bryson}, {Torres}, {Rowe},
  {Batalha}, {Borucki}, {Brown}, {Caldwell}, {Christiansen}, {Deming},
  {Fabrycky}, {Ford}, {Gilliland}, {Gillon}, {Haas}, {Jenkins}, {Kinemuchi},
  {Koch}, {Lissauer}, {Lucas}, {Mullally}, {MacQueen}, {Marcy}, {Sasselov},
  {Seager}, {Still}, {Tenenbaum}, {Uddin}, \& {Winn}}]{Desert-11}
{D{\'e}sert}, J.-M., {Charbonneau}, D., {Demory}, B.-O., {et~al.} 2011, \apjs,
  197, 14

\bibitem[{{Gibson} {et~al.}(2009){Gibson}, {Pollacco}, {Simpson}, {Barros},
  {Joshi}, {Todd}, {Keenan}, {Skillen}, {Benn}, {Christian}, {Hrudkov{\'a}}, \&
  {Steele}}]{Gibson-09}
{Gibson}, N.~P., {Pollacco}, D., {Simpson}, E.~K., {et~al.} 2009, \apj, 700,
  1078

\bibitem[{{Henry} {et~al.}(1997){Henry}, {Ianna}, {Kirkpatrick}, \&
  {Jahreiss}}]{Henry-97}
{Henry}, T.~J., {Ianna}, P.~A., {Kirkpatrick}, J.~D., \& {Jahreiss}, H. 1997,
  \aj, 114, 388

\bibitem[{{Kipping} \& {Bakos}(2011)}]{Kipping-11b}
{Kipping}, D. \& {Bakos}, G. 2011, \apj, 733, 36

\bibitem[{{Kipping}(2009)}]{Kipping-09}
{Kipping}, D.~M. 2009, \mnras, 392, 181

\bibitem[{{Mazeh} {et~al.}(2013){Mazeh}, {Nachmani}, {Holczer}, {Fabrycky},
  {Ford}, {Sanchis-Ojeda}, {Sokol}, {Rowe}, {Agol}, {Carter}, {Lissauer},
  {Quintana}, {Ragozzine}, {Steffen}, \& {Welsh}}]{Mazeh-13}
{Mazeh}, T., {Nachmani}, G., {Holczer}, T., {et~al.} 2013, ArXiv 1301.5499

\bibitem[{{O'Neal} {et~al.}(1998){O'Neal}, {Saar}, \& {Neff}}]{ONeal-98}
{O'Neal}, D., {Saar}, S.~M., \& {Neff}, J.~E. 1998, \apjl, 501, L73

\bibitem[{{Oshagh} {et~al.}(2013){Oshagh}, {Boisse}, {Bou{\'e}}, {Montalto},
  {Santos}, {Bonfils}, \& {Haghighipour}}]{Oshagh-12b}
{Oshagh}, M., {Boisse}, I., {Bou{\'e}}, G., {et~al.} 2013, \aap, 549, A35

\bibitem[{{Oshagh} {et~al.}(2012){Oshagh}, {Bou{\'e}}, {Haghighipour},
  {Montalto}, {Figueira}, \& {Santos}}]{Oshagh-12}
{Oshagh}, M., {Bou{\'e}}, G., {Haghighipour}, N., {et~al.} 2012, \aap, 540, A62

\bibitem[{{Pont} {et~al.}(2007){Pont}, {Gilliland}, {Moutou}, {Charbonneau},
  {Bouchy}, {Brown}, {Mayor}, {Queloz}, {Santos}, \& {Udry}}]{Pont-07}
{Pont}, F., {Gilliland}, R.~L., {Moutou}, C., {et~al.} 2007, \aap, 476, 1347

\bibitem[{{Pont} {et~al.}(2008){Pont}, {Knutson}, {Gilliland}, {Moutou}, \&
  {Charbonneau}}]{Pont-08}
{Pont}, F., {Knutson}, H., {Gilliland}, R.~L., {Moutou}, C., \& {Charbonneau},
  D. 2008, \mnras, 385, 109

\bibitem[{{Sanchis-Ojeda} \& {Winn}(2011)}]{Sanchis-Ojeda-11b}
{Sanchis-Ojeda}, R. \& {Winn}, J.~N. 2011, \apj, 743, 61

\bibitem[{{Sanchis-Ojeda} {et~al.}(2011){Sanchis-Ojeda}, {Winn}, {Holman},
  {Carter}, {Osip}, \& {Fuentes}}]{Sanchis-Ojeda-11a}
{Sanchis-Ojeda}, R., {Winn}, J.~N., {Holman}, M.~J., {et~al.} 2011, \apj, 733,
  127

\bibitem[{{Sing}(2010)}]{Sing-10}
{Sing}, D.~K. 2010, \aap, 510, A21

\bibitem[{{Solanki}(2003)}]{Solanki-03}
{Solanki}, S.~K. 2003, \aapr, 11, 153

\bibitem[{{Strassmeier}(1999)}]{Strassmeier-99}
{Strassmeier}, K.~G. 1999, \aap, 347, 225

\bibitem[{{Tas} \& {Evren}(2000)}]{Tas-00}
{Tas}, G. \& {Evren}, S. 2000, Information Bulletin on Variable Stars, 4992, 1

\end{thebibliography}

\end{document}